\documentclass[twocolumn,showpacs,preprintnumbers,amsmath,amssymb]{revtex4}

\usepackage{graphicx}
\usepackage{dcolumn}
\usepackage{bm}

\def\bm#1{\setbox0\hbox{$#1$}\kern+.03em\copy0\kern-\wd0
                             kern-.03em\copy0\kern-\wd0
                             \kern-.03em\box0\kern-.03em}
\mathchardef\nabla="7272
\def\pd{\partial}
\def\ie{{\it i.e., }}
\def\eg{{\it e.g., }}

\def\etal{{\it et al.}}
\def\viz{{\it viz., }}

\def\uu{{\bf u}}
\def\kk{{\bf k}}
\def\pp{{\bf p}}
\def\qq{{\bf q}}

\def\ee{{\bf e}}

\begin{document}

\title{Forced Stratified Turbulence: \\
Successive Transitions with Reynolds Number}

\author{J.-P. Laval}
 \altaffiliation[ ]{CNRS UMR 8107, LML, Blv. Paul Langevin, F-59655 Villeneuve d'Ascq, France}
\author{J. C. McWilliams}
 \altaffiliation[ ]{IGPP, UCLA, Los Angeles, CA 90095-1565, USA}
\author{B. Dubrulle}
 \altaffiliation[ ]{CNRS, URA 2464, GIT/SPEC/DRECAM/DSM, CEA Saclay, F-91191 Gif sur Yvette Cedex, France}

\date{\today} 


\begin{abstract}
Numerical simulations are made for forced turbulence at a sequence of
increasing values of Reynolds number, $R$, keeping fixed a strongly
stable, volume-mean density stratification.  At smaller values of $R$,
the turbulent velocity is mainly horizontal, and the momentum balance
is approximately cyclostrophic and hydrostatic.  This is a regime
dominated by so-called pancake vortices, with only a weak excitation
of internal gravity waves and large values of the local Richardson
number, $Ri$, everywhere.  At higher values of $R$ there are
successive transitions to (a) overturning motions with local reversals
in the density stratification and small or negative values of $Ri$;
(b) growth of a horizontally uniform vertical shear flow component;
and (c) growth of a large-scale vertical flow component.  Throughout
these transitions, pancake vortices continue to dominate the
large-scale part of the turbulence, and the gravity wave component
remains weak except at small scales.
\end{abstract}

\maketitle

\section{Introduction}
Most atmospheric and oceanic flows on intermediate scales of ${\cal
O}(1)-{\cal O}(10^3)$ m are strongly influenced by the stable vertical
density stratification, $d\overline{\rho}(z)/dz < 0$ (with $\hat {\bf
z}= \ee_3$ anti-parallel to gravity), but are less influenced by
Earth's rotation than flows on larger scales.  Consequently, the
turbulence in this regime is quite different from unstratified shear
turbulence, convection, and geostrophic turbulence, all of which have
been more extensively studied and whose behaviors are now relatively
familiar.  It has been argued \cite{lilly83} that the main effect of
strongly stable stratification --- \ie small Froude number, $Fr =
V/NL_v$, where $V$ is a horizontal velocity scale, $N = - (g/\rho_o)\,
d\overline{\rho}/dz$ is the Br\"unt V\"a\"\i ss\"ala frequency, and
$L_v$ is vertical length scale --- is to organize the flow into two
distinct, non-interacting classes: nearly linear internal gravity
waves and fully nonlinear stratified turbulence.  The flow patterns of
stratified turbulence are often called pancake vortices
\cite{kimura96} or vortical modes \cite{lelong91} because of their
small aspect ratio, $H/L$ (where $L$ is a horizontal length scale),
and significant vertical vorticity, neither of which is generally true
for internal gravity waves.  Pancake vortices have an anisotropic
velocity (primarily horizontal) and shear field (primarily vertical),
and they evolve principally under the nonlinear influence of
horizontal advection as in two-dimensional turbulence.  These motions
cause little vertical turbulent heat or mass flux, and they have a
highly anisotropic, inhomogeneous energy cascade to small scales and
dissipation \cite{godeferd94}.  At moderate values of Reynolds number
--- $R = VL_h/\nu$, where $\nu$ is the kinematic viscosity and $L_h$
is a horizontal length scale --- stratified turbulence evolves
self-consistently, at least in freely decaying flows, in the sense
that a bulk value for $Fr$ does not increase with time as energy
dissipation causes $R$ to decrease \cite{metais89}.  At leading order
in $Fr$, the inviscid dynamical balances for stratified turbulence are
equivalent to two-dimensional turbulence \cite{lilly83} evolving
independently at each vertical level.  The energy dissipation may be
modeled by adding a vertical eddy diffusion \cite{embid98} that acts
to couple vertically adjacent levels and diffusively selects a
limiting vertical length scale.  However, with the assumption of
uniform asymptotic validity as $Fr \rightarrow 0$, stratified
turbulence is constrained by hydrostatic and cyclostrophic force
balances that also act to couple adjacent layers and may internally
select a finite vertical scale as $R \rightarrow \infty$ without
inducing any vertically overturning motions \cite{mcwilliams85}.\

These behaviors have been studied both with laboratory experiments
\cite{yap93,fincham96} and with numerical simulations up to $R = {\cal
O}(100)$; a review of this subject has recently been made by Riley \&
Lelong \cite{riley00}.  In the ocean and atmosphere, $R$ values are
generally several orders of magnitude larger than those commonly
reached in experiments or numerical simulations.  Thus, an important
open question is whether the preceding wave-turbulence partition
remains valid at very large $R$.  A central part of this question is
whether the pancake vortices persist and remain ``stable'' with
respect to overturning motions.\

The dynamical stability properties of a stably stratified shear flow
are usually related to the Richardson number, $Ri$, locally
defined by
\begin{equation}
Ri \ = \ -\frac{g}{\rho_o} \, \frac{\pd{\rho}}{\pd z} \,
        \left(\frac{\pd{\bf u}_h}{\pd z}\right)^{-2} \
     = \ \left(N^2 + \frac{\pd{\theta}}{\pd z}\right) \,
        \left(\frac{\pd{\bf u}_h}{\pd z}\right)^{-2},
\label{eq:richardson}
\end{equation}
where ${\bf u}_h$ is the horizontal velocity and $\theta = - g
\rho'/\rho_o$ is the normalized ``temperature'' associated with density
fluctuations $\rho'$.  Alternatively, $Ri$ is defined as a bulk
measure in an analogous fashion with $N^2$ in the numerator and shear
variance in the denominator (\ie a bulk $Ri \sim Fr^{-2}$).  In the
inviscid limit, a sufficient condition for stability of a parallel
vertically sheared flow (\ie a Kelvin-Helmholtz flow) is that the
local $Ri$ exceed 0.25 throughout the flow \cite{miles61,howard61}.
Gage \cite{gage71} refined this criterion for several simple viscid
shear flows and obtained values of the critical $Ri$ for linear
instability between about 0.05 and 0.11 for large $R$.  In a numerical
simulation at very high resolution (\ie with a maximum Reynolds number
based on the shear layer thickness $Re_ H = 24000$) and moderate
stratification, Werne \& Fritts \cite{werne99} show that the
turbulence organizes itself so that $Ri$ never exceeds 0.25.  In the
more complex natural environment, velocity variance and estimates of
the vertical mixing efficiency increase rapidly as $Ri$ decreases
through the range between about 0.5 and 0.0 \cite{peters88}.  On the
other hand, Majda \& Shefter \cite{majda98} stress the importance of
temporal behavior on flow stability by constructing a family of time
periodic solutions that are unstable at arbitrarily large $Ri$. \

In this paper we examine the behavior of stratified turbulence at
large $R$ values by means of numerical simulations of the Boussinesq
Equations with forcing at the larger scales of the computational
domain.  Some simulations of forced stratified turbulence have
been performed previously by Herring and M\'etais \cite{herring89},
but the available resolution did not allow clear conclusions for flows
at high $R$.  Most experiments and simulations for stratified
turbulence have been conducted on decaying turbulence, with many
focused specifically on the transition from isotropic to stratified
turbulence after an initial high-energy excitation event
\cite{metais89,staquet98,fincham96}.  Because of the large energy
dissipation rate in both isotropic and stratified turbulence, this
evolutionary path starts with large $R$ and $Fr$ and only briefly
resides in a fully developed regime {\it en route} to small $R$ and
$Fr$; this situation therefore provides a limited view of the
equilibrium geophysical regime.  One of our main purposes here is to
analyze the flow regimes of stratified turbulence in terms of $R$ and
$Fr$ varied independently in a controlled fashion.

\section{Description of the Calculation}

\subsection{Governing Equations}

A pseudo-spectral code is used to integrate the Boussinesq Equations
on a triply periodic domain \cite{vincent91}; \viz
\begin{equation}
\left \{
\begin{array}{lll}
\partial_t {\bf u}  + {\bf u} \cdot \nabla  {\bf u}    &=&
 - \nabla p + \theta \;{\ee_3} +  \nu \Delta \uu  + {\bf F}, \\
\partial_t \theta + \uu \cdot \nabla \theta &=&
 - N^2  {\bf u} \cdot {\ee_3} + \kappa \Delta \theta,  \\
\nabla \cdot  {\bf u} &=& 0.
\end{array}
\right.
\label{eq:ns3dstr}
\end{equation}
In these equations, ${\bf u}$ is the three-dimensional velocity, $p$
is the pressure divided by $\rho_o$, $\kappa$ is the conductivity, and
${\bf F}$ is the imposed forcing.  In Fourier space the equations for
the Fourier components of velocity and the temperature,
$\hat{\uu}(\kk)$ and $\Theta(\kk)$, are
\begin{equation}
\left \{
\begin{array}{l}
\left(\partial_t  + \nu  k^{2} \right) \hat{u}_i(\kk,t) 
- P_{i3}(\kk) \, \Theta(\kk,t) \\
= -i \, k_l P_{in}(\kk) \, \int_{\kk+\pp+\qq=0} \hat{u}_n(\pp,t) 
\hat{u}_l(\qq,t) \, d^3\pp + \hat{F_i}(\kk,t), \\
\\
\left(\partial_t  + \kappa k^{2} \right) \Theta(\kk,t) 
+ N^2 \, \hat{u}_3(\kk,t) \\
= -i \, k_n \int_{\kk+\pp+\qq=0} \hat{u}_n(\pp,t) 
\hat{\Theta}(\qq,t) \, d^3\pp . 
\end{array}
\right.
\label{eq:boussinesqk}
\end{equation}
$P_{ij}=\delta_{ij} - k_ik_j/k^2$ is the projection operator onto
the plane orthogonal to $\hat{\uu}(\kk)$; $\delta_{ij}$ is the
Kronecker tensor, repeated indices indicate summation; and $i^2 = -1$
when $i$ does not appear as an index.

\subsection{Energy Decomposition}

As a simple means of separating turbulence ({\it v}ortices) and {\it
w}aves, as well as a horizontally uniform {\it s}hear flow that is
neither of these, we use Craya's decomposition \cite{craya58} for the
incompressible velocity field in Fourier space into orthogonal
components $\hat{\uu}_v$, $\hat{\uu}_w$, and $\hat{\uu}_s$:
\begin{equation}
\begin{array}{llll}
\hat{\uu}(\kk,t)  &=&   \hat{\uu}_v(\kk,t) + \hat{\uu}_{w}(\kk,t) & 
        \mbox{\hspace{5mm} if} \;\; \kk_h \neq 0, \\
\hat{\uu}(\kk,t)  &=&   \hat{\uu}_s(\kk,t)  & 
        \mbox{\hspace{5mm} if} \;\; \kk_h  = 0, 
\end{array}
\end{equation}
where
\begin{eqnarray}
\hat{\uu}_v(\kk,t) = \hat{\phi}_v(\kk,t) \; \varphi_v(\kk), \\
\hat{\uu}_w(\kk,t) = \hat{\phi}_w(\kk,t) \; \varphi_w(\kk)
\end{eqnarray}
and 
\begin{eqnarray}
\varphi_v(\kk) &=& [ (\kk \times \ee_3) ] / \vert (\kk \times \ee_3) \vert, \\
\varphi_w(\kk) &=& [ \kk \times (\kk \times \ee_3))] / \vert \kk \times 
(\kk \times \ee_3) \vert.
\end{eqnarray}
$\kk_3 = \kk \cdot \ee_3$ and $\kk_h = \kk - \kk_3$ are the
components of the wavenumber perpendicular and parallel to gravity.
The components $(\hat{\phi}_v,\hat{\phi}_w)$ were previously used by,
\eg Riley \etal~\cite{riley00} and Lilly \cite{lilly65}, and they are
usually refered to ``vortical'' and ``wave'' components.  The
emergence of the ``shear'' component $\hat{\uu}_{s}$ was emphasized by
Smith and Waleffe~\cite{smith02}.  Associated with $\hat{\phi}_v$,
$\hat{\phi}_w$ and $\hat{\uu}_s$, we define the kinetic energy spectra
by
\begin{eqnarray}
\Phi_{v}(k) &=& \frac{1}{2}  \sum_\pp 
        \hat{\phi}^\ast_{v}(\pp) \hat{\phi}_{v}(\pp), \\
\Phi_{w}(k) &=& \frac{1}{2}  \sum_ {\pp, \pp_h \ne 0} 
        \hat{\phi}^\ast_{w}(\pp) \hat{\phi}_{w}(\pp), \\
\Phi_{s}(k) &=& \frac{1}{2}  \sum_{\pp, \pp_h = 0} 
        \hat{\uu}^\ast_{s}(\pp) \hat{\uu}_{s}(\pp),
\end{eqnarray}
where the sum $\sum$ is done over a shell ${k-1/2 <\vert \pp \vert <k+1/2}$.
We further define the ``available potential energy'' spectrum  by
\begin{equation}
\Phi_p(k) = \frac{1}{2} \sum_\pp \frac{\hat{\Theta}^\ast(\pp) 
        \hat{\Theta}(\pp)}{N^2} ,
\end{equation}
and the ``total kinetic energy'' spectrum by
\begin{equation}
\Phi(k) = \frac{1}{2}  \sum_\pp 
        \hat{u}^\ast(\pp) \hat{u}(\pp) .
\end{equation}
Total component energies ($E_{v}$, $E_{w}$, $E_{s}$, $E_{p}$, and $E$)
are obtained by summing over all shells $k$.  In addition to this
decomposition, we define the ``vertical kinetic energy'' $E_{z}$ as
half the area-averaged square of the vertical velocity, $\uu \cdot
\ee_3$, and $\Phi_z$ as the vertical kinetic energy spectrum.

\subsection{Posing the Problem}

Our analysis is based on a single numerical solution (designed after
calculating many more exploratory solutions than we wish to admit).
An anisotropic grid resolution is used to take advantage of the
anisotropy of the flow that arises in response to the stable
stratification (\ie a small aspect ratio).  The calculation is performed
over a vertical fraction of a cubic domain by imposing a $2
\pi / M$ vertical periodicity of the velocity ($M = 8$ in the present
case).  For a given number of grid points, this increases the
achievable $R$ value without loss of generality as long as the typical
vertical scale is much smaller than the horizontal periodicity length.
The level of stratification is controlled by adjusting the spatially
uniform value of N in time.  The forcing {\bf F} is defined by ${\bf
F}(\kk,t) \, dt = \beta(k,t) \, {\bf u}(\kk,t)$, with $\beta$ chosen
so that the difference $\Phi_f(k,t^o)$ of the energy spectra before
and after the forcing (\ie the energy injection rate) is constant in
time:
\begin{eqnarray}
\Phi_f(k,t^o) &=& \frac{1}{2} \sum_{\pp} \left\{ \vert \hat{\uu}(\pp,t) 
        + dt \, \hat{F}(\pp,t)\vert^2 \right\} \ - \ \Phi(k,t) \nonumber \\
        &=& \left( 1+\beta(k,t) \right)^2 \; \Phi(k,t)  .
\label{eq:force1}
\end{eqnarray}
The coefficient $\beta(k,t)$ is obtained from eq. (\ref{eq:force1}) as
\begin{equation}
\beta(k,t)\  = \ \sqrt{\frac{\Phi(k,t) + \Phi_f(k,t^o)}{\Phi(k,t)}} \ - \ 1 .
\end{equation}
We choose to force only the first vertical and horizontal modes (\ie
$\Phi_f(k,t^o)\ne 0$ for $k=k_{v}^o=M$ and $k=k_{h}^o=1$).  In order
to reach a high enough $R$ value, a hyper-diffusion ($(-1)^p \nu_p
\Delta_h^p$, with p=4 and a small coefficient of $\nu_p=10^{-12}$) is
added to the Newtonian diffusion in the horizontal direction.  Several
additional simulations at higher horizontal resolution demonstrate
that this modification does not qualitatively affect the results
presented here.  Since most of the dissipation occurs in association
with shear in the vertical direction (about 99.5\% in the stable
pancake regime; see below), we use ordinary Newtonian diffusion in
this direction to have a clean definition of Taylor's Reynolds number,
$R_\lambda = U_{rms} \lambda / (\epsilon / D) = 2 \sqrt{D} E /
\epsilon$, where $U_{rms} = \sqrt{2E}$ is the RMS velocity, $D$ is the 
enstrophy (\ie vorticity variance), $\lambda = \sqrt{2E/D}$ is the
Taylor scale, and $\epsilon$ the dissipation rate).  The Prandtl
number $\nu/\kappa$ is set to 1.  The bulk Froude number is defined by
$Fr_v= U_{rms}/ (H N)$, where $H=3\pi/4E (\sum_{k_3} k_3^{-1}
\Phi_v(k_3))$ is the typical vertical scale.\

\begin{figure} 
   \includegraphics[width=\columnwidth]{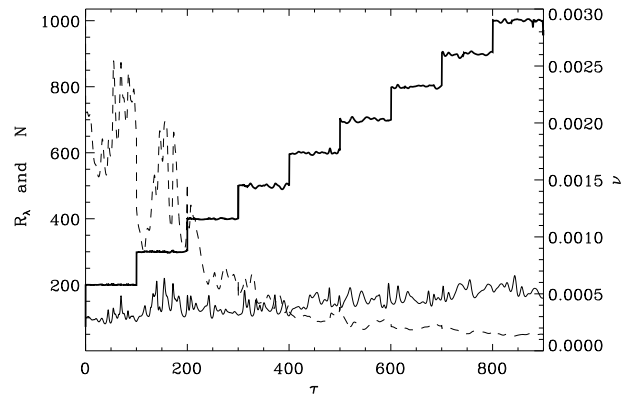}
   \caption{\label{fig:n-nu} Experimental path for the primary
    simulation: the time evolution 
   of Taylor's Reynolds number $R_\lambda$ (thick line), the resulting
   Br\"unt V\"a\"\i ss\"ala frequency N (thin line), and the viscosity
   $\nu$ (dashed line).  N and $\nu$ are adjusted to maintain the Froude
   number at the constant value of 0.08 and make $R_\lambda$ follow the
   indicated history.}
   \end{figure}

The primary simulation is adjusted in time to follow a given
experimental path in terms of $R_\lambda$ and $Fr_v$ by adjustment in
time of $N$ and $\nu$.  The evolution of these two parameters is shown
in Fig. \ref{fig:n-nu}. 
$Fr_v$ is maintained at a small value of 0.08, and $R_{\lambda}$ is
changed from 200 to 1000 by steps of 100 after integration periods of
100 turnover times (\ie $T=L/U_{rms}$ with L the integral scale
defined by $L=3\pi/4E (\sum_{k} k^{-1} \Phi(k))$).  This period is
approximately long enough to adjust to an equilibrium state for each
$R_{\lambda}$.  This experimental path is designed to expose a
sequence of regime transitions with increasing $R$.\

The equations are solved on a $2 \pi \times 2 \pi \times 2 \pi/M $
domain with a spatial grid size of $256 \times 256 \times 64$ for
$R_{\lambda} < 500$ and $256 \times 256 \times 128$ for higher
$R_{\lambda}$.  The simulation appear to be slightly under-resolved
for the highest Reynolds number in the sense that the dissipation
range is not extensive in $\Phi(k)$.  The ratios of the highest
resolved ($k^m_h=96$ and $k^m_v=341$) to the corresponding Kolmogorov
wavenumbers (${k^{\eta}_h}=(\nu^3/\epsilon_h)^{-1/4}$ and
${k^{\eta}_v}=(\nu^3/\epsilon_v)^{-1/4}$) are 1.1 and 0.92,
respectively, for $R_\lambda = 500$, and they are 0.4 and 0.2 for
$R_\lambda = 1000$.  The under-resolution in the horizontal direction
is compensated by the hyper-viscosity.  In the vertical direction, the
kinetic energy at the resolution scale is much smaller than at the
largest scales in the dissipation range (by a factor of $10^{-5}$ at
the highest $R_\lambda$). The Ozmidoz scale ($Lo=(\epsilon/N^3)^{1/2} \simeq 4.10^{-4}$ for $R_\lambda = 700$) is more than one order of magnitude smaller than the smallest resolved vertical scales. This ratio means that all the resolved scales are significantly influenced by stratification.\

The simulations were made on one processor
of a NEC SX-5.  The code uses optimized ASL libraries of Fast Fourier
Transforms and requires approximately 4 seconds per time step at the
highest resolution.  The time step is chosen to sufficiently resolve
the fastest wave oscillations with period $2\pi / N$.  The simulation
is integrated over a total of 360,000 time steps and takes about
200 hours.

\section{Solution Analysis}

\begin{figure} 
   \includegraphics[width=\columnwidth]{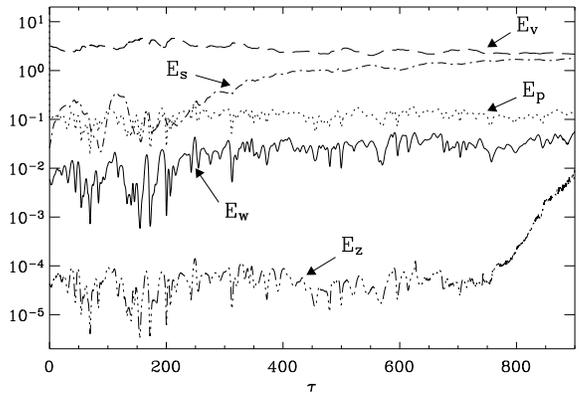}
   \caption{\label{fig:evol_energy} History of the energy
     components in the primary simulation. }
   \end{figure}

The experimental path for the primary solution is demonstrated in
Fig. \ref{fig:n-nu} as a function of the non-dimensional time,
$\tau=t/T$, normalized by the turnover time $T$.  The corresponding
histories of the energy components are shown in
Fig. \ref{fig:evol_energy}.  As $R_\lambda$ increases, we see a
sequence of regime transitions.\

\begin{figure} 
   \includegraphics[width=\columnwidth]{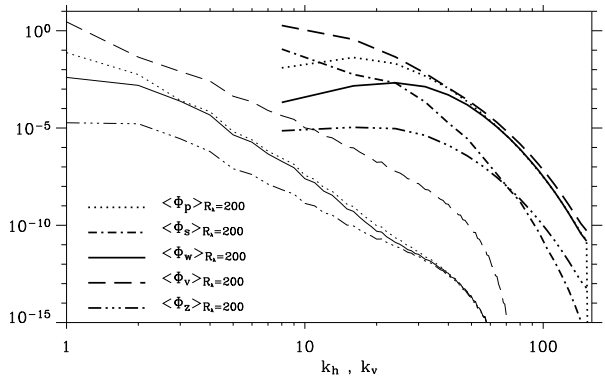}
   \caption{\label{fig:spectra0} Energy spectra with respect
     to horizontal and vertical wavenumbers averaged over more than
     100 times at $R_\lambda = 200$ for the primary
     simulation.}
   \end{figure}

\begin{figure} 
   \includegraphics[width=\columnwidth]{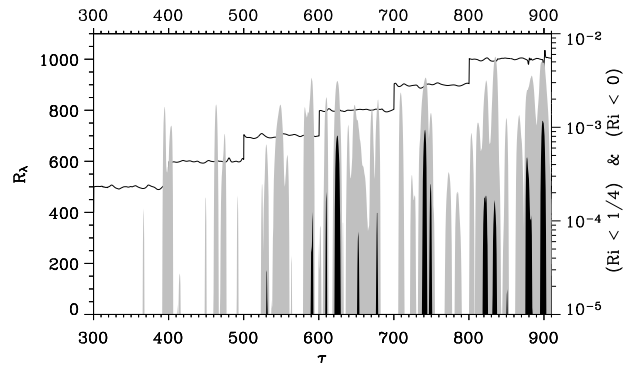}
   \caption{\label{fig:evol_local_ri_000_025} Time evolution of $R_\lambda$
   (solid line) and the volume fraction  of the 
   domain with local $Ri < 0.25$
   (filled grey area) and with local $Ri < 0$ (filled black area).  
   There was no occurrence  of $Ri < 0.25$ for $0 < \tau < 300$. }
   \end{figure}

At moderate $R_\lambda$, less than $\sim 400$, the pancake motions are
stable (\ie the local $Ri$ is everywhere large), and the vortical
energy dominates all other energy components at all wavenumbers
(Fig. \ref{fig:spectra0}).\
The first transition occurs for $R_\lambda \approx 400$, and it is
evident in the significant growth of energy in the shear component,
$E_{s}$ (Fig. \ref{fig:evol_energy}).
The next transition, for $R_\lambda \approx 500$, is evident in the
intermittent occurrence of regions with small local values of $Ri$
below the Kelvin-Helmholtz critical inviscid stability value of 0.25
(Fig. \ref{fig:evol_local_ri_000_025}).
A further transition, for $R_\lambda \approx 700$, is evident in local
violations of the inviscid gravitational stability critical value of
$Ri = 0.0$ (Fig. \ref{fig:evol_local_ri_000_025}).  Finally, we see
yet another transition for $R_\lambda \approx 900$, evident in the
growth of vertical kinetic energy, $E_{z}$
(Fig. \ref{fig:evol_energy}).  Interestingly, throughout all these
transitions, the principal measures of the internal-wave energy,
$E_{w}$ and $E_{p}$, show little change relative to the vortical-mode
energy, $E_{v}$.  Since $E_{v}$ itself remains reasonably constant
with time and its horizontal spectrum $\Phi_{v}(k_h)$ maintains a
similar shape and magnitude at low wavenumbers, we conclude that
pancake motions are indeed persistent throughout the $R_\lambda$ range
we have been able to explore, even though the structure and intensity
of the flow changes substantially at high wavenumbers, in the
horizontally uniform vertical shear, and in the vertical velocity.

\subsection{Growth of the Vertical Shear Component}

\begin{figure} 
   \includegraphics[width=\columnwidth]{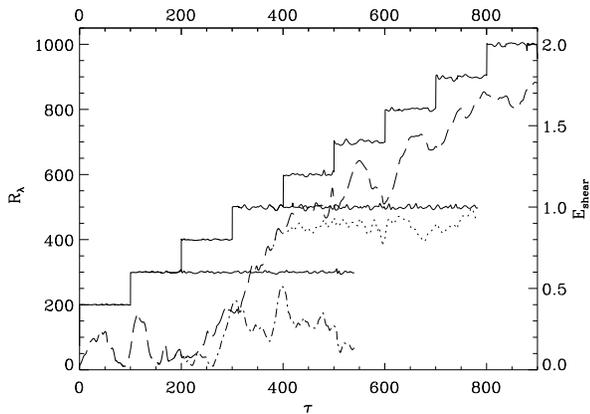}
   \caption{\label{fig:comp_shear_energy} History of $R_\lambda(t)$ 
   (solid lines) and $E_{s}(t)$ for three simulations. 
    The primary simulation (with a dashed line for
   its $E_{s}$) has an increasing Reynolds number by steps of 100 from
   200 to 1000, and the second and third simulations have constant
   Reynolds numbers of $R_{\lambda}=300$ (dash-dotted line) and
   $R_{\lambda}=500$ (dotted line), starting from the primary simulation
   at $\tau=200$ and $\tau=400$, respectively.}
   \end{figure}

In Fig. \ref{fig:evol_energy} the vortical energy is nearly steady
over the entire simulation.  The wave energy is more variable, but on
average it is steady as $R_\lambda$ increases.  However, the shear
energy is a growing function of time.  It represents an inverse
horizontal cascade of kinetic energy into $k_h = 0$.  The intensity of
the inverse cascade of shear energy is probably a function of the
location of the energy peak in the horizontal direction; in the
present case, the forcing is imposed at $k_h = 1$, and a substantial
part of the energy is transformed into pure vertical shear.  To assess
the degree of equilibration for this inverse cascade, two additional
simulations are made.  Both start from the primary simulation and
thereafter hold $R_\lambda$ constant for several hundred turnover
times, but their starting $(\tau, \ R_\lambda)$ values differ.
Fig. \ref{fig:comp_shear_energy} shows that $E_{s}$ does indeed
equilibrate over a period of less than 100 turnover times at a level
that increases systematically with $R_\lambda$.\

This growth of shear kinetic energy has been seen previously when the
Froude number is below a critical value \cite{smith02}.  For the
alternative Froude number defined by $Fr_s= (\epsilon \,
{k_v^o}^2)^{1/3}/N)$, our simulations have a value of approximately
0.025, more than an order of magnitude below the identified critical
value of 0.42.  In this previous study, the shear kinetic energy did
not equilibrate even after more than 1000 turnover times.  This may be
due to its reliance on hyper-diffusion in all directions that exerts
only a weak damping on the shear component.

\subsection{Onset of Overturning Motions}

\begin{figure} 
   \includegraphics[width=\columnwidth]{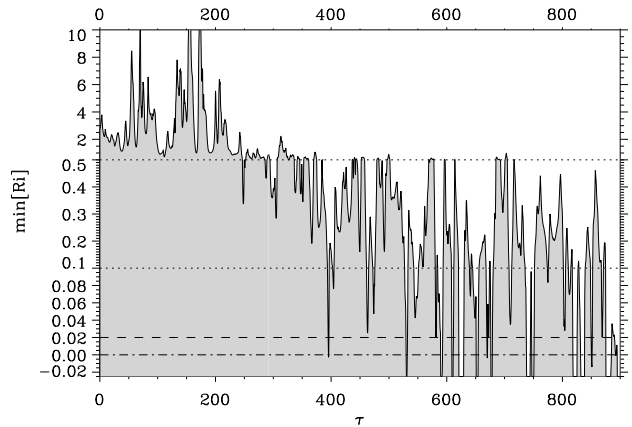}
   \caption{\label{fig:evol_min_ri} History of the global minimum value 
    of local $Ri$ for the primary simulation.  The ordinate scale is split
    into three intervals to see both large and small values on the same plot.
     Whenever the minimum $Ri$ drops below about 0.02, it continues to develop
    into overturning in the density profile (\ie min[Ri] $<$ 0); this first 
    occurs around $\tau
   \simeq 400$.}
   \end{figure}

Overturning occurs when an unstable shear layer rolls up, pulling
high-density fluid above low-density fluid.  This instability can
occur in a fairly localized way.  Its occurrence is detected with the
local $Ri$ defined in eq. (\ref{eq:richardson}), and an indication of
the overturning instability is a negative $Ri$ value.  Because of the
Miles-Howard condition, we can expect that this Kelvin-Helmholtz
instability only occurs in regions where $Ri$ is initially less than
$0.25$.  The history of the fraction of the domain with the local $Ri$
below 0.25 is shown in Fig. \ref{fig:evol_local_ri_000_025}.  More
events with $Ri < 0.25$ happen as $R_\lambda$ increases, although they
evidently remain intermittent.  The regime of overturning events first
appears at $R_\lambda=700$ ($\tau \simeq 535$).  Even at the highest
$R$ value, only a small fraction of the domain is actively overturning
at any time, less than 0.7\%.  The Probability Density Function (PDF)
for $Ri$ shows that most of the domain remains far away from
overturning, with only a small tail in the PDF that extends to small
and negative $Ri$ values.  Because of the intermittency, long
averaging periods are required for stable statistics.  Nevertheless,
we can say that there is a well determined value of $R$ for the onset
of overturning events.  To demonstrate this, the regime just before
the first overturning event in the primary simulation
($R_\lambda=500$) is integrated over a longer period (250 turnover
times), and no overturning occurs despite several events with $Ri <
0.25$.  To characterize more precisely the critical $Ri$ for
instability, we follow the history of the global minimum in local $Ri$
(Fig. \ref{fig:evol_min_ri}).
The minimum $Ri$ value that does not immediately lead to local
overturning is $Ri \simeq 0.02$.  This value is a bit smaller but of
the same order of magnitude than the values computed by Gage
\cite{gage71} for simple shear flows.  \

\begin{figure} 
   \includegraphics[width=0.8\columnwidth]{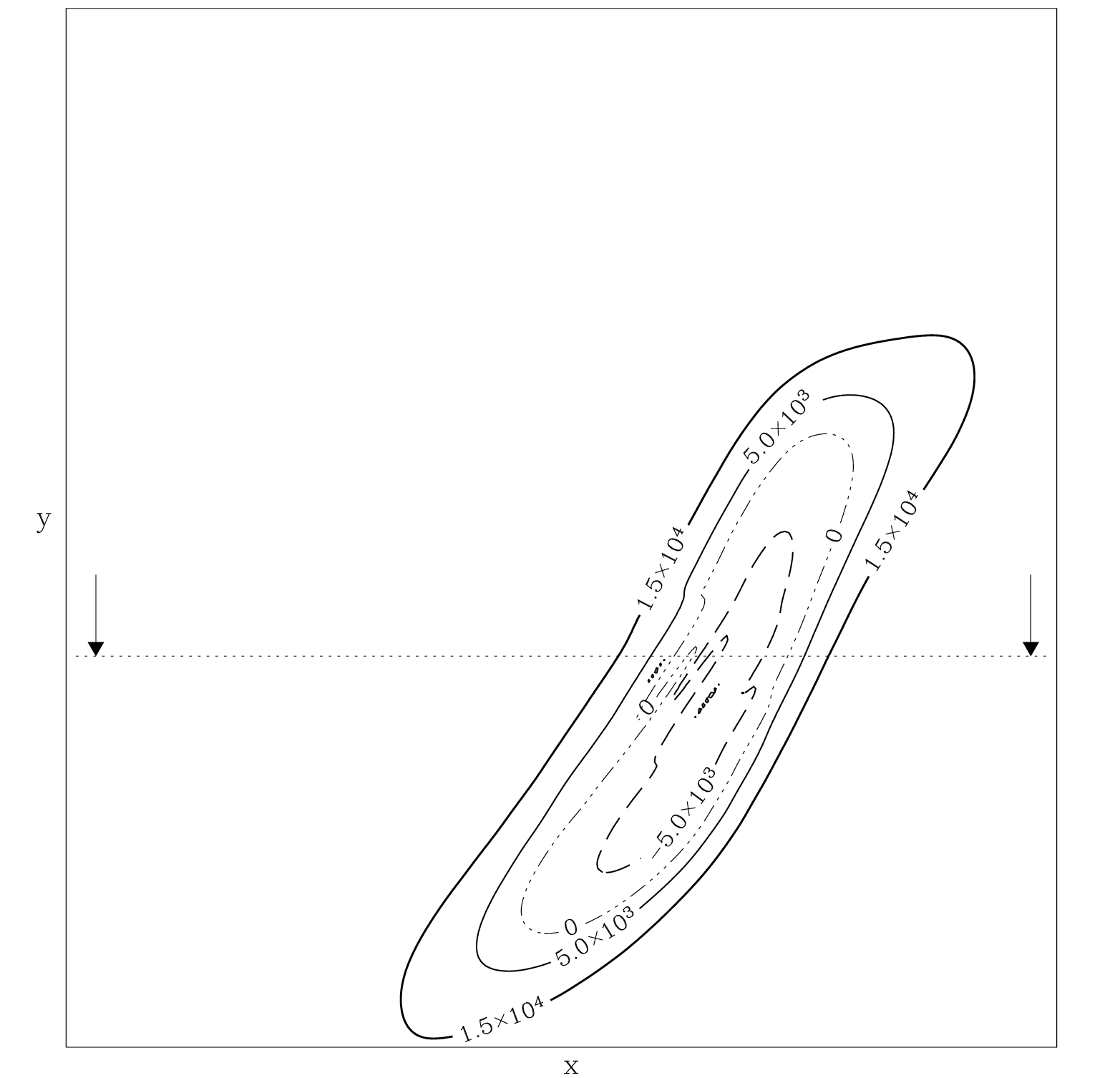}
   \includegraphics[width=0.8\columnwidth]{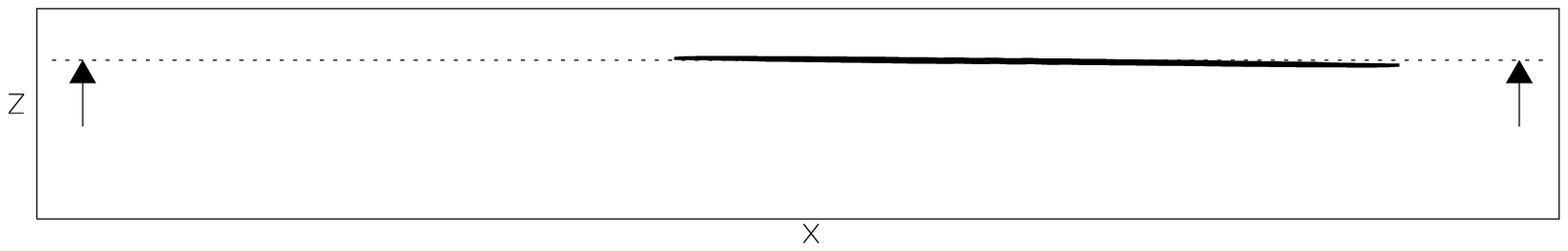}
   \caption{\label{fig:slice_vdg}Instantaneous isolines of
    vertical density gradient at
    $\tau = 741$ ($R_\lambda = 900$) for two perpendicular slices 
     of the whole domain  at $z=cte$ (upper graph) and $y=cte$ (lower graph) .}
   \end{figure}

\begin{figure} 
   \includegraphics[width=\columnwidth]{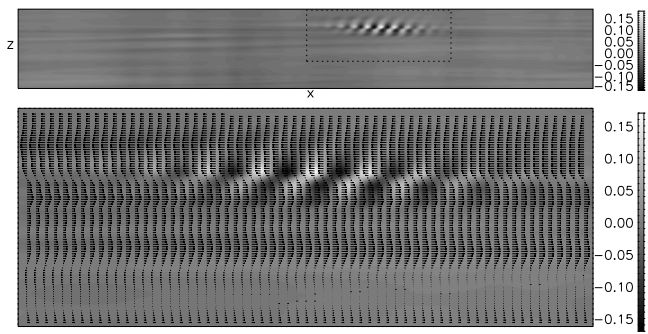}
   \includegraphics[width=\columnwidth]{figure-08b.ps}
   \caption{\label{fig:slice_vz} Vertical slice (same than in Fig. \ref{fig:slice_vdg})
   of vertical velocity at $\tau = 741$ ($R_\lambda = 900$) (upper graph) and a zoom
    of the region of intense vertical velocity with the projection of velocity
   vectors (lower graph) .}
   \end{figure}

The spatial distribution of small $Ri$ events is organized into thin
sheets with large vertical shear in the horizontal velocity.  An
example of a region with a negative vertical density gradient is shown
Fig. \ref{fig:slice_vdg}.
The vertical size of this region is very thin (a few grid cells) even
though its horizontal size at this time ($\simeq 2 \pi / 3$) is not
much smaller than to the domain size.  An intense vertical velocity is
associated with domain of negative vertical density gradient
(Fig. \ref{fig:slice_vz}), but the present simulation has only a
marginally adequate resolution to expose the convective overturning
events.\

\begin{figure} 
   \includegraphics[width=0.8\columnwidth]{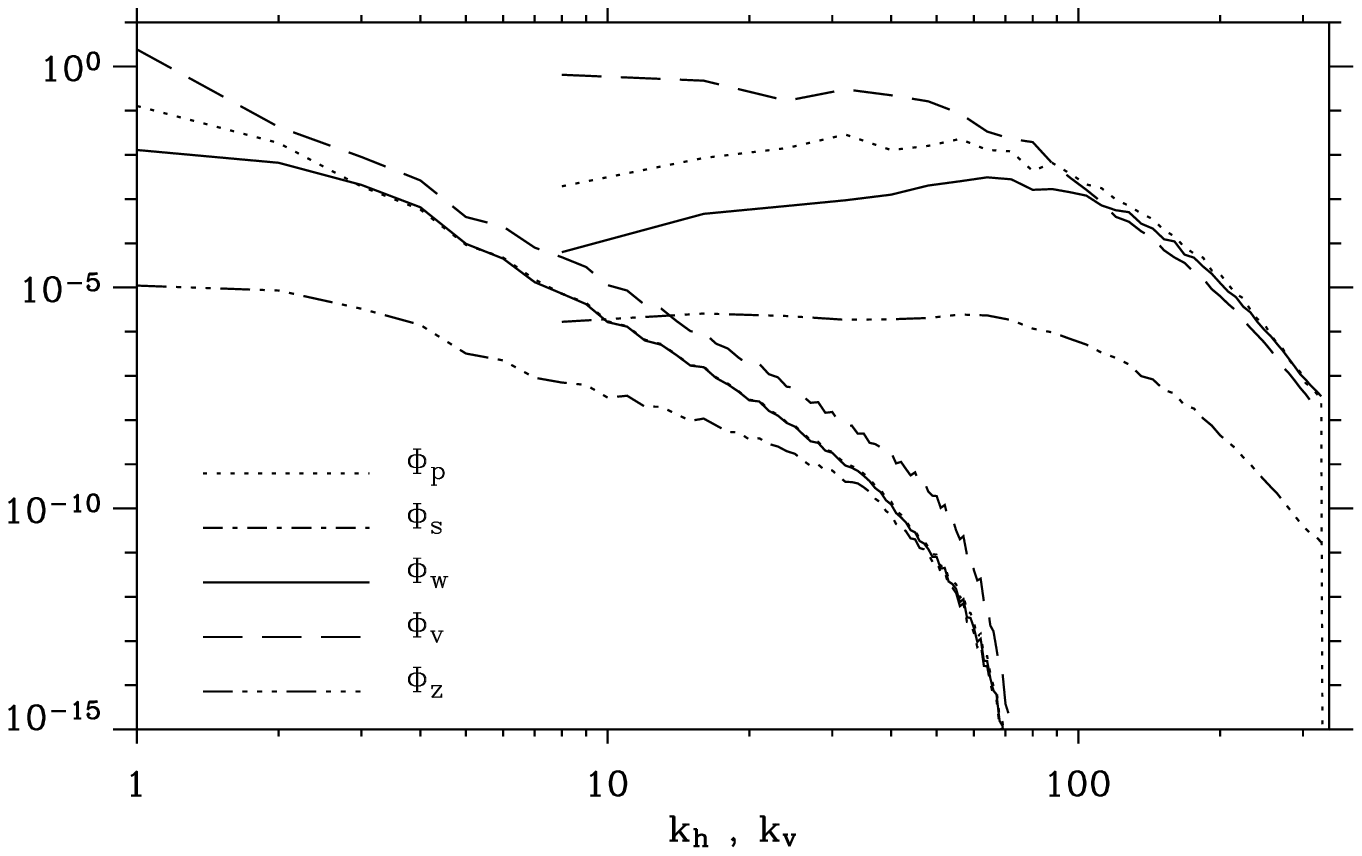}
   \includegraphics[width=0.8\columnwidth]{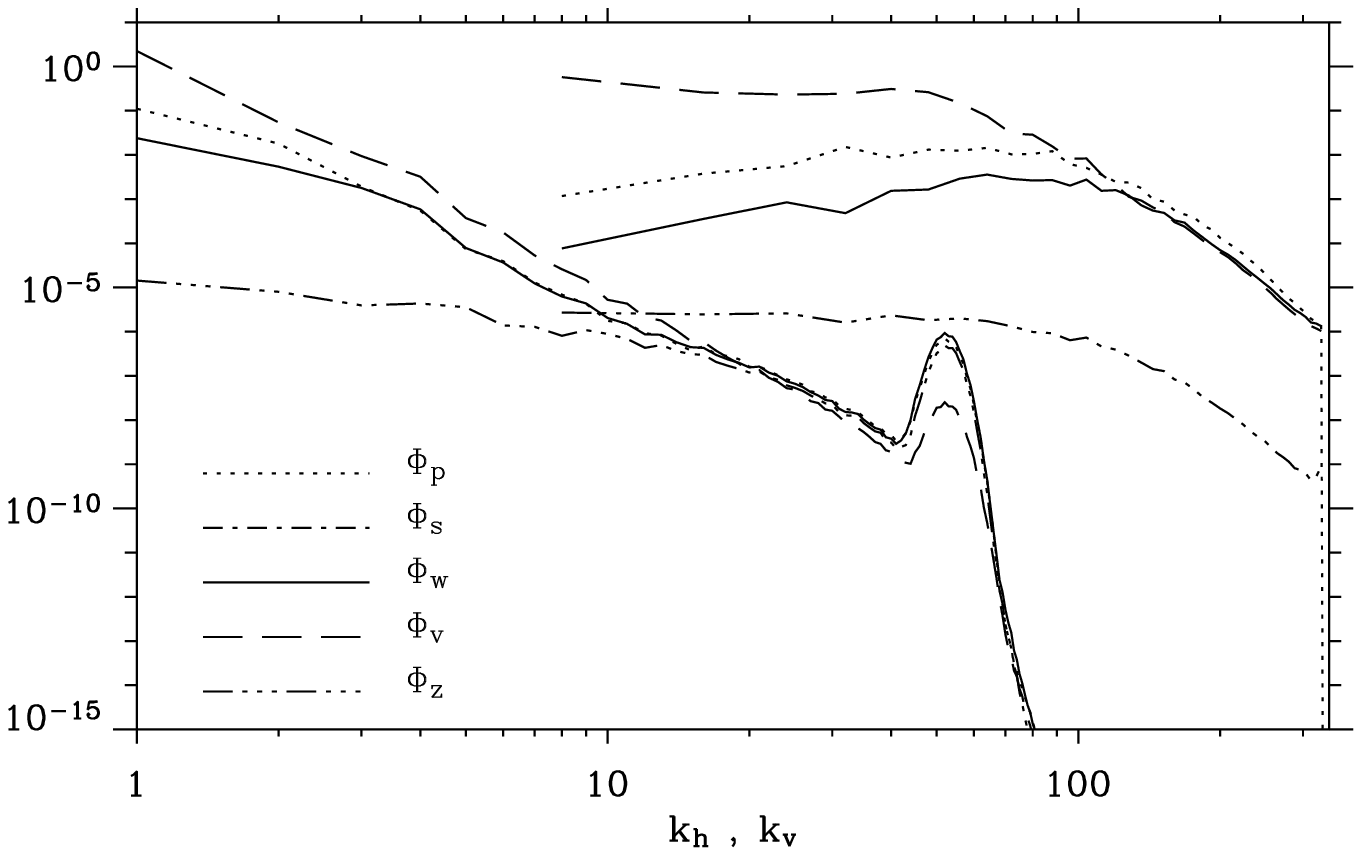}
   \caption{\label{fig:spectra12}Energy spectra with respect to horizontal
    and vertical wavenumbers at $\tau=521$ ($R_\lambda = 700$) (upper graph)
    and $\tau=741$ ($R_\lambda = 900$) (lower graph) for the
    primary simulation. The local energy peaks at  $\tau=741$ for $40 < k_h < 60$ are an
   indication of one or several overturning events($Min[Ri]>0$).}
   \end{figure}

Fig. \ref{fig:spectra12} shows the spectra for the energy components 
at $\tau=521$, before any overturning occurs.  The spectra with
respect to horizontal wavenumber are very steep, with a slope close to
$k_h^{-5}$ for $\Phi_v$, and the shape does not vary much with
$R_\lambda$ in this stably stratified regime without overturning.  The
onset of the overturning instability does not seem to affect
significantly the overall level of wave and vortical energy
(Fig. \ref{fig:evol_energy}).  However, the overturning events are
easily identifiable as a peak in the horizontal spectra at $\tau=741$
(Fig. \ref{fig:spectra12}).  The peak is located at the horizontal
scale $L_h \simeq 2 \pi / 50$ that matches the typical vertical
scale, $L_v$.  At this particular time, the instability is localized
in a single region of the domain.\

\begin{figure} 
   \includegraphics[width=0.8\columnwidth]{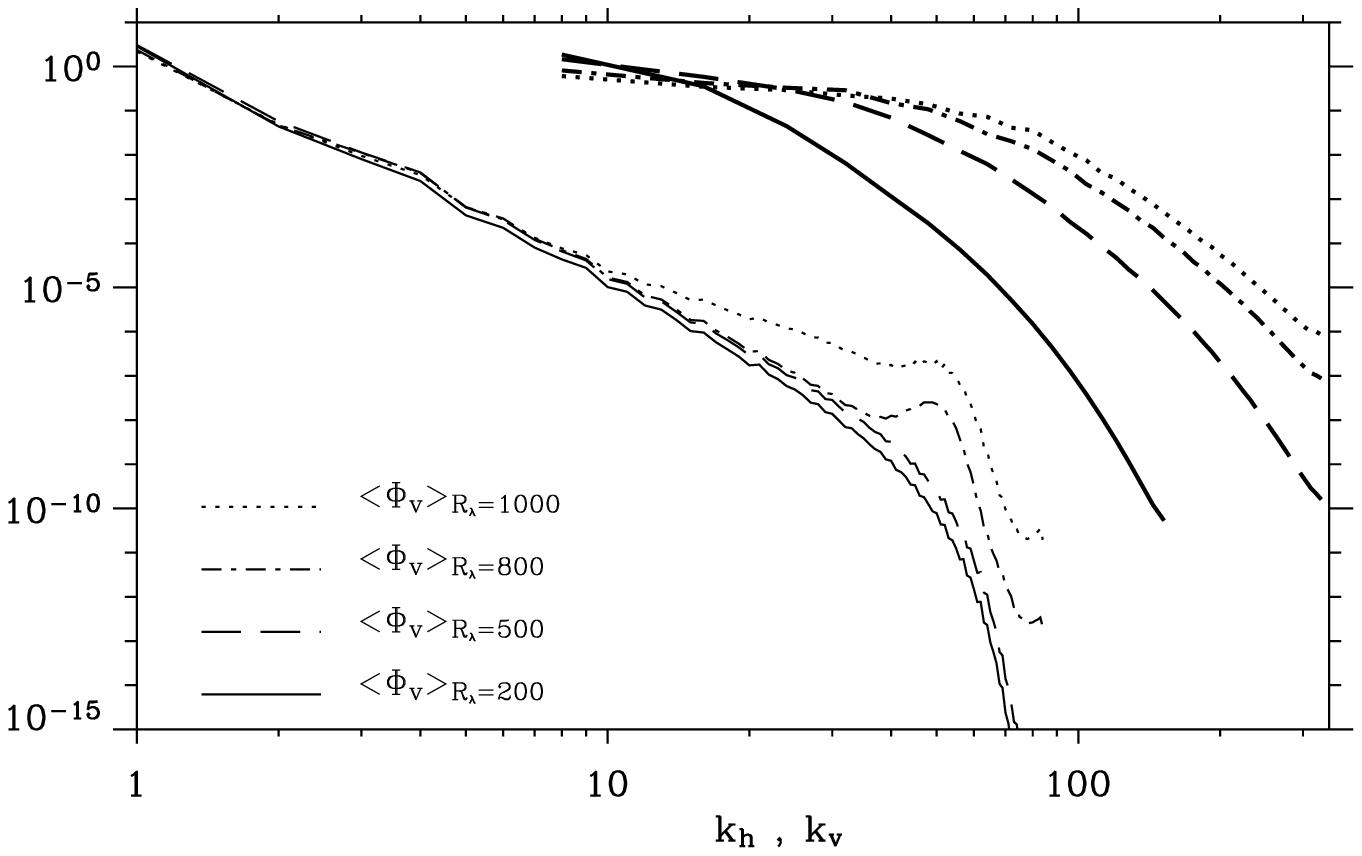}
   \includegraphics[width=0.8\columnwidth]{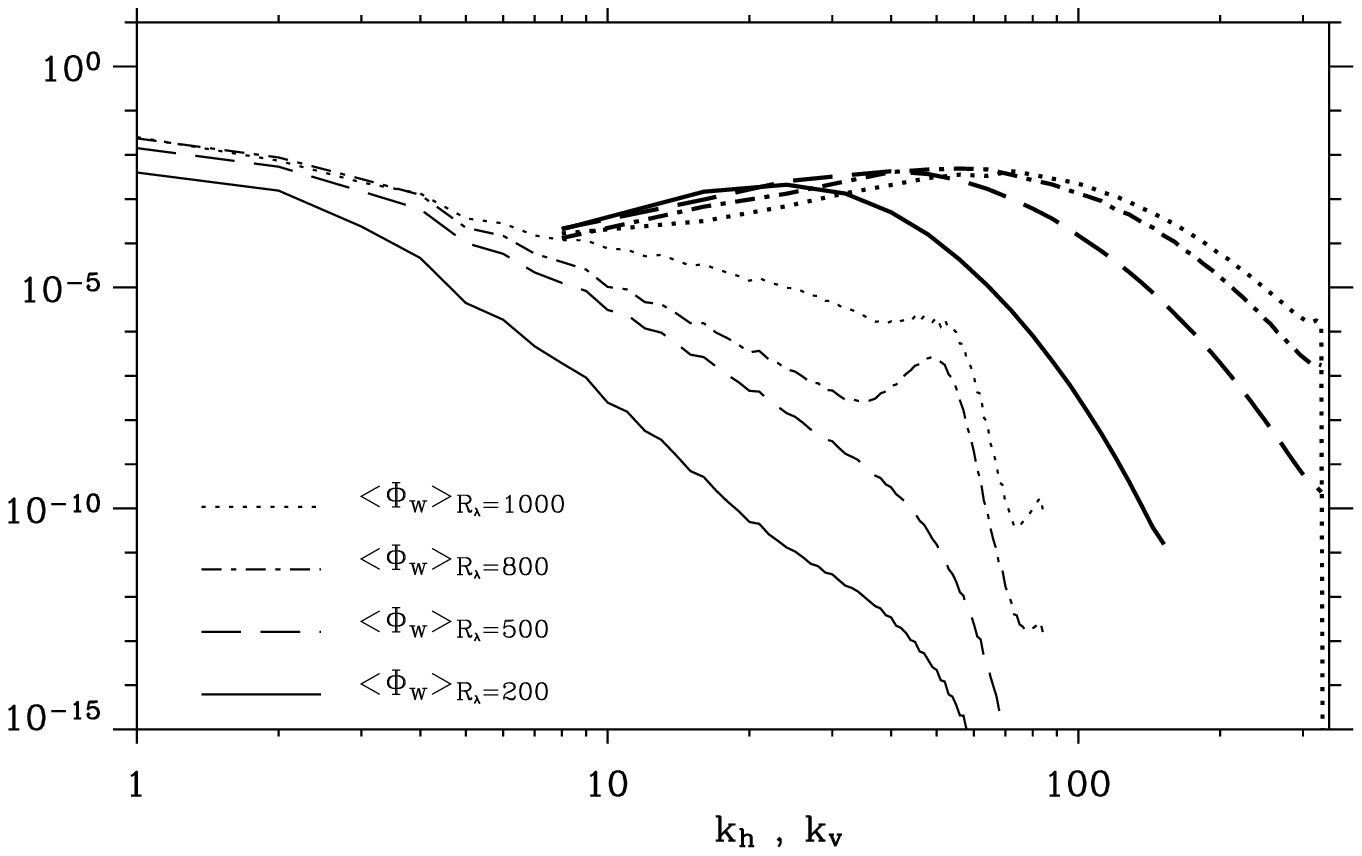}
   \caption{\label{fig:comp_spect_av_new}Comparison of vortical (upper graph)
    and wave (lower graph) energy spectra with respect to 
   horizontal and vertical wavenumbers for four different Reynolds
   number.  Each spectrum is an average over more than 200
   times for each $R_\lambda$.  The horizontal spectra are very similar
   at large horizontal scales but differ at small scales ($k_h > 10$).
   The situation is different in the vertical spectra, where the typical
   scale decreases with $R_\lambda$.}
   \end{figure}

The energy at the largest horizontal scales is not affected by the
overturning instabilities that are localized in both space and time.
In Fig. \ref{fig:comp_spect_av_new},
time-averaged vortical-energy spectra are compared for four values of
$R_\lambda$.  The horizontal spectra are very similar up to the
typical scales of the overturnings.  However, at both finer horizontal
scales and at all vertical scales finer than the forcing scale, the
spectrum amplitude increases systematically with $R_\lambda$.  At the
constant, small $Fr$ value in this simulation, the vertical spectrum
slope becomes quite shallow as $R_\lambda$ increases.

\subsection{Growth of Large-Scale Vertical Motions}

\begin{figure} 
   \includegraphics[width=\columnwidth]{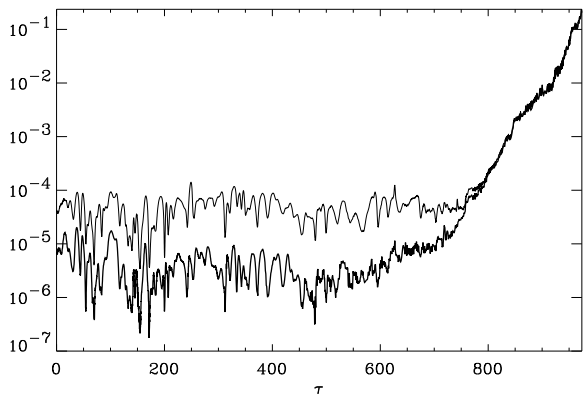}
   \caption{\label{fig:vertical_energy} Evolution in time of the total 
   vertical energy (thin line) and the vertical energy at the zero vertical
    mode (thick line) for the primary simulation (parameters of Fig. \ref{fig:n-nu}). The
    component of the zero vertical mode is several order of magnitude
    smaller than the total vertical energy for $\tau < 700$ and become
    the dominant mode for $\tau > 800$}
   \end{figure}

The energy histories in Fig. \ref{fig:evol_energy} expose another
transition at an even larger $R_\lambda \simeq 900$, \viz the systematic
growth of vertical kinetic energy $E_z$.  Inspection of the vertical
energy spectrum reveals that the growth of $E_z$ after $\tau=700$
occurs principally at $k_v = 0$ (Fig. \ref{fig:vertical_energy}).
This mode of instability is reminiscent of the ``negative-viscosity
instability'' observed in a Kolmogorov flow \cite{dubrulle91b},
further investigated by Dubrulle \& Frisch \cite{dubrulle91a} with a
multi-scale analysis.  This analysis shows that a parallel flow with a
small transverse scale develops a negative-viscosity instability to
large-scale perturbations in the transverse direction when the
viscosity becomes less than the RMS value of the streamfunction of the
primary flow.  In our simulation, it is difficult to test precisely
this criterion of instability because the streamfunction is
ill-defined.  Crudely, we can expect an instability of this type when
$\nu < \Phi_{s}(\ell_v)^{1/2} \ell_v$, where $\ell_v$ is the scale of
the vertical shear and $\Phi_{s}(\ell_v)$ is the shear energy at this
scale.  In our simulation this relation is satisfied on average for
$\tau>700$ if $2\pi / \ell_v > 80$.  This scale is comparable in
magnitude with the typical vertical wavenumber $2 \pi / L_v \simeq
50$.  However, due to the complexity of the forced stratified flow, it
is difficult to prove the nature of this instability pending
more apt stability analyses.

\section{Summary and Discussion}

In our simulations of forced, equilibrium, stratified turbulence, we
see behaviors somewhat different from many previous studies of
decaying stratified turbulence that were not able to sustain a large
value of the Reynolds number, $R$.  Most often the criteria for the
occurrence of pancake vortices and suppression of overturning motions
(\ie Kelvin-Helmholtz and gravitational stability) have been linked to
the stratification $N$ but rarely to $R$.  Indeed, we find that the
stability of a solution is mainly controlled by two parameters with
opposite effects on stability: increasing N (decreasing $Fr$) leads to
a more stable solution and decreasing $\nu$ (increasing $R$) has the
opposite effect.  For a fixed low value of $Fr$, we follow an
experimental path of increasing $R$ far enough to detect several
regime transitions beyond the familiar one of stable pancake vortices.
One transition is to the intermittent occurrence of regions with small
or negative $Ri$.  This refutes previous arguments
\cite{lilly83,mcwilliams85} that stratified turbulence remains stable
with uniformly small local values of $Fr$ at large $R$ and with
uniformly cyclostrophic, hydrostatic diagnostic momentum balances.
This transition may plausibly be associated with inviscid
Kelvin-Helmholtz and ensuing gravitational instabilities of the
pancake vortices, although in our simulations the viscous effects on
the unstable scales are significant.  Nevertheless, the pancake
vortices continue to be the energetically dominant component of the
turbulence even up to the highest $R$ values examined here, and
visualizations of the large-scale potential vorticity field (not
shown) show little change in spatial structure with $R_\lambda$.\

Two other transitions to large-scale motions other than pancake
vortices do occur: a first one to growth of the shear kinetic energy
at zero horizontal wavenumber and a second one to growth of vertical
kinetic energy with its spectrum peak at zero vertical wavenumber for
large $R$.  The former has been seen previously in stratified
turbulence \cite{smith02}, and the latter may be associated with
negative-viscosity instability seen previously in unstratified shear
flow \cite{dubrulle91b}.  Each of these large-scale transitions may be
interpreted as an inverse energy cascade.  However, they behavior is
strongly constrained by the domain size in our simulations where the
forcing is imposed at the gravest finite wavenumbers.  We will explore
beyond this limitation in future reports.\

In this paper we choose to focus on simulations at very small Froude
number, and we are able to reach a Reynolds number high enough to
destabilize the pancake vortices in several ways.  This leads us to
advance the following proposition about the nature of equilibrium
stratified turbulence: for any Froude number, no matter how small,
there are Reynolds numbers large enough so that a sequence of
transitions to non-pancake motions will always occur, and, conversely,
for any Reynolds number, no matter how large, there are Froude numbers
small enough so that these transitions are suppressed.  Obviously,
this hypothesis warrants further testing, as do our provisional
interpretations of the dynamical nature of the transitions.\

\section*{Acknowledgments}
The primary simulation was calculated on the NEC SX-5 of the {\it
Institut du D\'eveloppement et des Ressources en Informatique
Scientifique (IDRIS)} and the IBM RS6000/SP of the {\it Centre de
Ressource Informatique} of the University of Lille was used for
additional simulations.  JPL and JCM acknowledge support from the
Office of Naval Research (grant N00014-98-1-0165).

\end{document}